\documentclass[pra,aps,groupedaddress,twocolumn,floatfix,nofootinbib]{revtex4-2}
\usepackage{graphicx}
\usepackage{soul}
\usepackage{float}
\usepackage{centernot}

\newcommand{\beq}{\begin{equation}}
\newcommand{\eeq}{\end{equation}}
\newcommand{\beqa}{\begin{eqnarray}}
\newcommand{\eeqa}{\end{eqnarray}}

\def\half{\frac{1}{2}}
\def\opone{\leavevmode\hbox{\small1\normalsize\kern-.33em1}}
\def\gbar{\bar g}

\begin{document}

\title{From Quantum Cryptography to Intuitionism and beyond:\\relativity, many-worlds and non-locality}
\author{Nicolas Gisin \\
\it \small Group of Applied Physics, University of Geneva, 1211 Geneva 4,    Switzerland\\
Constructor University, Bremen, Germany}

\date{\small \today}
\begin{abstract}
After a recollection of some of my encounters with Gilles Brassard, on the occasion of his 70th birthday I present, first, a new idea on creation of new information at a limited rate. Next, I develop the idea that ``indeterminacy is relative" [Entropy {\bf23} 1326 (2021)] and how this illuminates the compatibility between relativity and non-locality.
\end{abstract}
\maketitle

\section{Introduction}\label{intro}
When I first met Gilles Brassard, in 1993 at a workshop in Dagstuhl-Germany, I was already the older one of the two of us. But due to my highly non-linear career\footnote{In the late 1980's and early 1990's I was a telecom engineer working in the promising field of optical communications, optical fibers and the burgeoning internet.}, Gilles was by far the most prominent scientist. Another explanation of this fact is that Gilles started science when most kids were still playing with puppets. Anyway, we quickly became friends. I was so happy and proud when I could explain him how to perform unambiguous quantum measurements \cite{unambiguousMeas} and that I knew how to demonstrate quantum key distribution in standard optical fibers \cite{MullerQKDNature95}.

Years later Gilles became a ``many-worldist", while I remained a ``collapse person". This did not harm our friendship, though we clearly acknowledged our different opinions. I develop on this in section \ref{RelInd}.

My second intellectual encounter with Gilles is much more recent.
In 2019 I published in a philosophy journal a paper entitled ``Indeterminism in Physics, Classical Chaos and Bohmian Mechanics: Are Real Numbers Really Real?" \cite{NGrealNb}. This was the start for me of an entirely new research program. I realized that many of the quantum features -- those often presented as paradoxes -- were actually already present in indeterministic classical physics \cite{del2024features}. For instance, the need of an interpretation is equally present in classical mechanics: if one rejects infinite information density in space (or in phase space), then one has to conclude that the usual real numbers are not physical\footnote{As anticipated by Max Born who wrote: ``Statements like `a quantity $x$ has a completely definite value' (expressed by a real number and represented by a point in the mathematical continuum) seem to me to have no physical meaning" \cite{born}.} and that classical mechanics is, as quantum mechanics according to standard interpretations, an indeterministic theory. And then, once one contemplates the possibility of indeterministic classical mechanics, one is tempted to search for additional hidden variables, a feature I used to think typical of quantum theory. Doing so, one quickly realizes that the classical hidden variables are but the standard real numbers, with their infinite list of inaccessible digits \cite{NGHiddenReals}.

Once one understands that ``real numbers are not really real" -- that they are not physical -- comes the question by what to replace them. Initially, with my close collaborator Flavio Del Santo, we invented something we named ``Finite Information Quantities", (FIQs) \cite{FlavioNG19}. Nice, but not real mathematics \cite{callegaro2020comment,del2020reply}. During a workshop in Jerusalem that I co-organised with my friend Yuval Dolev, I presented our ideas. After my talk, a local professor, Carl Posy, with a kippah on his head, came to me and said: ``you are reinventing intuitionism, we should have lunch together". Hence, I met intuitionism \cite{PosyBook,Bentzen}, a form of constructive mathematics in which time is essential \cite{StandfordEncyclodediaIntuitionism}. Somehow, intuitionistic real numbers do not come all at once, with all their infinite series of digits, but digits come into existence one after the other, as time passes\footnote{Brouwer, the father of intuitionism, thought that only rational numbers exist as a completed object. However, in more modern terms it is natural to consider that all computable numbers exist as completed objects. Indeed, by definition computable numbers are entirely contained in a (finite) algorithm. Here, when I write ``intuitionistic real numbers", I mean typical real numbers, those that are not computable.}. Fascinating! Here is a well developed mathematical language in which time plays a fundamental role. With such a language, I thought, it should be possible to do physics in a way that would allow us to speak of time not merely as parameter, but as a real process.

I wrote yet another paper for a philosophy journal \cite{gisin2021indeterminismSynthese} (and a second one for physicists \cite{NGNaturePhysComment20}), explaining that the mathematical language we speak when doing physics has a huge influence on the world-view that physics presents us. I illustrated this with classical Platonic time-less mathematics and with intuitionistic mathematics, the first one leading naturally to a deterministic world-view, while the second one provides an indeterministic world-view. I sent that preprint to Gilles who immediately replied ``oh yes, I know intuitionism, I even wrote notes on it!".

This was my second deep encounter with Gilles, the first one being quantum cryptography. But with Gilles things are never simple.

\section{Intuitionism, digits and information}
Gilles likes to be precise, very precise\footnote{During our first email exchanges, in French, Gilles used to correct my spelling mistakes! I loved it.}. In my presentations I often illustrate intuitionistic mathematical numbers as numbers where the digits do not appear all at once -- as they do in classical Platonic mathematics -- but instead appear one after the other, as in a process that occurs as time processes. This provides the correct intuition. Important is that at any finite time, only finite information exists. But this finite information grows as time processes. Somehow, classical mathematics is intuitionistic mathematics seen, so to speak, from the end of time, once everything settled to well determinate values and nothing new happens.

This nicely illustrates, I believe, both the finiteness of information and the importance of time in intuitionism. However, this is not very precise. More precisely, as time passes new bits of information are created and these new bits could well affect more than the last digit. Actually, it could change all the digits, as Gilles started to explain to me per emails.

As usual Gilles took the question very seriously. This resulted in a nice publication by Gilles and colleagues in which they explained that I was not the first one committing the mistake of confusing new digits and new information \cite{BrassardDruckproblem}: Alan Turing made the same mistake \cite{TuringMistake}! Hence, I found myself in excellent company.

In intuitionistic mathematics the law of the excluded middle does not hold. The next digits -- sorry, the next bits of information -- are not yet determinate, hence propositions about their values have, presently, no truth value. I love this because it perfectly reflects the physical fact that future quantum measurement outcomes have, presently, no values. Similarly, if one interprets classical mechanics as an indeterministic theory, then there too the future is open. For example, the temperature in Geneva in precisely 10 years from now has, presently, no value. I believe that all this is pretty obvious, although it contradicts standard classical physics, is in big tension with classical timeless mathematics and is even impossible to formulate according to the classical logic we all have learned. Indeed, classical logic asserts both the law of the excluded middle -- all propositions have a truth value, either true or false -- and that the truth value of a proposition never changes: if true at some time $t$, then it has always been true \cite{creativeTime,
logic4Physicists}. Hence, since the temperature in 10 years will be determinate, this temperature, according to classical logic, has to have already now and in all past that same determinate value.

This is not the place to present intuitionism and indeterministic classical physics in any detail (see, e.g.~\cite{PosyBook,gisin2021indeterminismSynthese,FlavioNG19}). I believe that anyone who has not been too much infected by Laplace determinism, anyone who ``understands" that the future is open, that things can really happen, that new bits can come into existence, all those people feel that something is rotten in today's physics. Admittedly, quantum theory allows for randomness. Actually, both Heisenberg uncertainty -- better named indeterminacy -- principle and the violation of Bell inequalities ``prove" the existence of randomness (``prove" in quotation marks because all proofs require some assumptions, those needed to conclude randomness based on the violation of Bell inequalities are extremely weak, almost impossible to reject) \cite{Pironio10,Acin16}. But quantum randomness is too often seen as a challenge, not as the fuel of time, see section \ref{creatTime}.

Before proceeding I feel the need to distinguish myself from standard intuitionism in two ways. First, intuitionism is usually entangled with idealism (Brouwer was almost a solipsist!). I am a realist: things exist out there independent of me and of any agent. Information is not limited to information of agents, but represents a useful measure of structures existing out there. Hence, new information can be created out there without the need of any agency. With Bruno Bentzen and Flavio Del Santo we wrote a note on this entitled {\it Naturalistic Intuitionism} \cite{NatInt}. Second, intuitionistic logic is consistent and very nice. In particular it adequately does not assume the law of the excluded middle: some propositions have, presently, no value. However, once they gain a value, this never changes. A sort of intuitionistic logic with some time parameter would be even closer to the need of indeterministic physics \cite{PosyLogic4IndPhys,logic4Physicists}.

There remains, however, one delicate question: when do these indeterminate quantities (e.g.~temperature) gain determinacy? The temperature here in Geneva in 10 years is presently indeterminate -- ontologically indeterminate, not merely unknown or uncertain: even nature doesn't know. However, in ten years this temperature will be determinate. But when precisely does it gain a determinate value? This could be a kind of continuous process, the temperature in 10 years gains more and more a precise value, though it always remains with some indeterminacy, eventually a very small indeterminacy. Note that the same holds, I believe, for the position of the centre of mass of a billiard ball: its position is very precise, but not with infinite precision. Let's come back to the example of the temperature. As time passes, its value gains determinacy. How and when does that happen? A first possible answer inspired by Copenhagen-like interpretations goes as follows: when some agent performs a measurements and if this measurement is more precise than the determinacy of the measured quantity, then the quantity -- temperature in our example -- has to gain determinacy. Ok, but that can't be the full story. First because temperatures existed long before agents and long before the invention of thermometers. Next, because we don't really know what is a measurement apparatus. This is the well-known measurement problem: how can a photon or an atom know that it should stop obeying the Schr\"odinger equation -- or Hamilton's equation if it is a classical particle -- just because it encounters a bunch of atoms with a sticker saying ``measurement apparatus" \cite{QMsticker}! In short, in all models of indeterministic physics, we face a measurement problem. More precisely, we face the question of when do physical quantities gain determinacy. Or equivalently, when do potentialities actualize \cite{FlavioNG19,del2024features}?

I am convinced that this measurement problem, in particular the quantum measurement problem, is a real physics problem. It is not merely a problem of logic or something requiring some clever little twists here and there. It is really physics in the sense that the solution(s) will lead to new physics, with new predictions and new technologies. I do not know the solution to this fascinating problem. However, in honour of Gilles I dare to present in the next section a very bold idea. I am confident Gilles will like it -- maybe not the idea, but the fact that I present it in his honour.

\section{Finite information density in space and in time}\label{creatTime}
Let's start with classical mechanics interpreted as an indeterministic theory where, at all times, all quantities have finite information. For example, in this model all molecules have positions and momenta with some small indeterminacy. In a gas, the evolution is chaotic, hence the initial small indeterminacy becomes large. The phase-space volume remains constant, but the shape of an initially quasi-point-like bubble extends fingers\footnote{Not arbitrarily, a premise of Heisenberg indeterminacy principle is already present, see \cite{noArbitraryFingers}.} such that quantities like temperature (mean square of velocities) become impossibly indeterminate. Something has to happen. Recall the classical answer: the initial bubble is infinitely narrow, the initial condition is infinitely precise, represented by a mathematical point in phase space\footnote{This is clearly a hidden variable as there is no way to access this mathematical point \cite{NGHiddenReals}.}. Hence, classically one speaks of deterministic chaos: it is all determinate since the very beginning, since the big-bang, nothing really happens, it's only as in a movie (except that there are no spectators, hence no emotions, no feeling of time).

\begin{figure}[h]
\includegraphics[width=8cm]{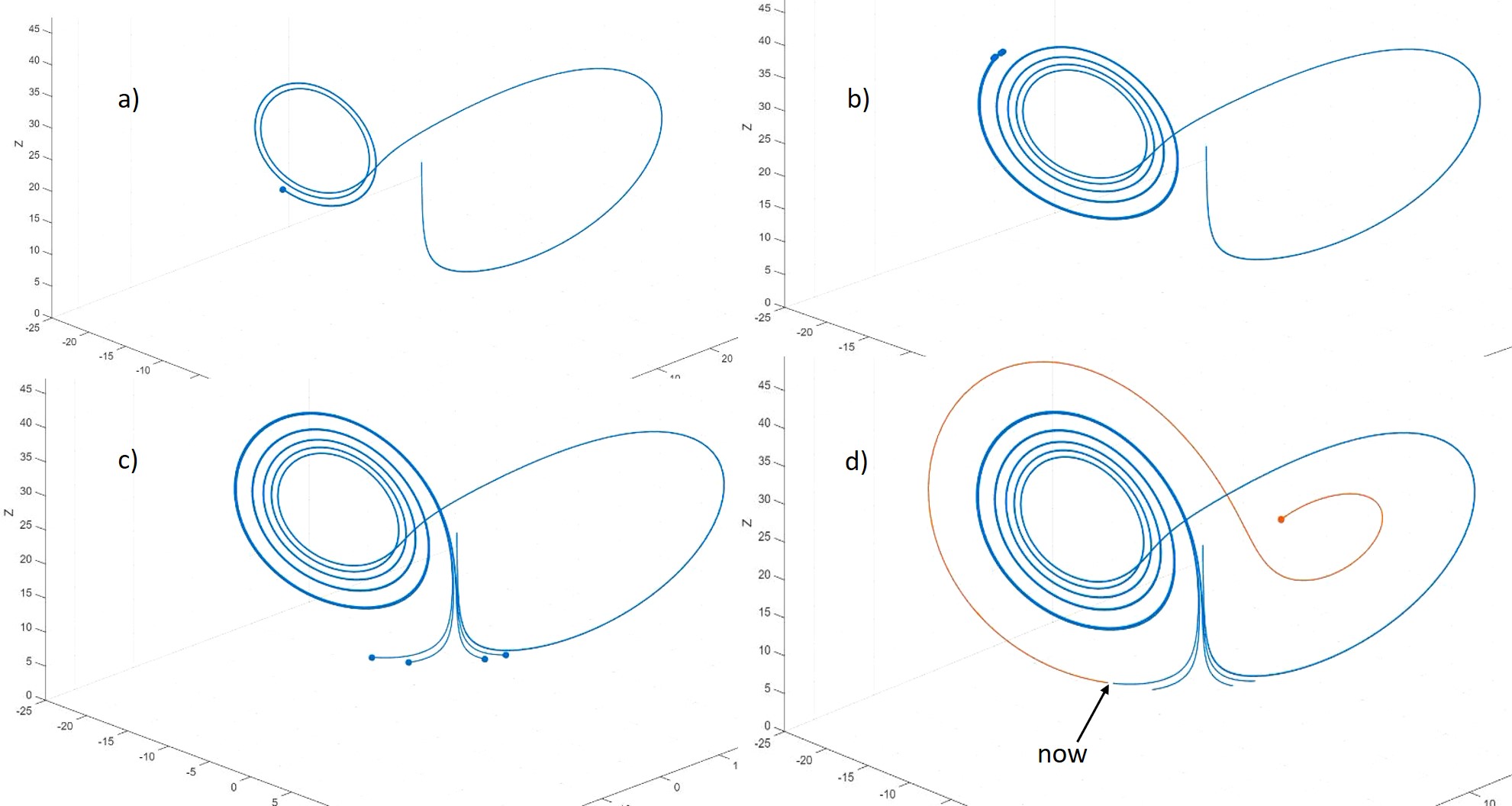}
\caption{\it Simulation of the 3-dimensional Lorenz chaotic dynamical system \cite{LorenzChaos}. a) The simulation starts somewhere near the center of the figure; let's assume the left wing represents a hot day, while the right one represents a cold day. On this simulation, the temperature quickly cooled down, but then went to the left, i.e.~the temperature increased and became warm. b) Actually, the initial condition was represented by 4 points, initially all within a pixel; but quickly, thanks to the high sensitivity of chaotic systems, the 4 points are visible. c) After yet some more time, the simulation shows a ``classical beam-splitter": some points go to the left (hot day), others to the right (cold day). d) As hot and cold can't co-exist, nature has to make up her mind; in this case she ``chooses" for a hot day, only one point on the left survives, illustrating a ``classical collapse", better said an act of actualization that indicates the now. The simulation is due to Mr Sajjad Bagdadi.}
\end{figure} 

Let's come back to indeterministic finite information classical physics. Something has to happen, but what? In Fig.~1 the well-known Lorenz chaotic system is displayed \cite{LorenzChaos}. Admittedly, this system is not Hamiltonian, but nicely uses only 3 dimensions, making illustrations easier; note that all classical chaotic dynamical systems have similar features to those illustrated here. 

Initially the state indeterminacy is within a pixel, Fig.~1.a, but after some finite time, the different points inside that pixel get split: some move to the right, some to the left, as in a (classical) beam-splitter, Fig.~1.c. In Fig.~1, in order to make the simulation easier and more visible, the initial small bubble is represented by only 4 points (clearly visible in Fig.~1.c, where 2 points move to the left and 2 to the right). However, it is straightforward to extend this picture to a full bubble. Assume the left wing of the trajectory represents warm weather, while the right hand side represents cold temperatures. Both cannot happen simultaneously. Nature has to make up her mind: either cold or hot, not both. In Fig.~1 chance decided for a warm day. But that was not predetermined. Randomness was at work. 

I like the story above very much. But that's not the end. Time continues to process and the ``point" -- small bubble -- on which, by chance, the system ``collapsed" (a classical collapse as you noticed) will again extend fingers and nature will again have to make up choices. But how precisely is the state determinate after a collapse? Nothing in indeterministic classical mechanics answers that question. A natural guess, I believe, would be to claim that there is a minimal size in phase-space, let's say $\hbar$ per degree of freedom. This is obviously inspired by quantum mechanics.

Let me be explicit, the goal is not to regress to classical mechanics, quantum mechanics is a better theory, there is no way back. However, by considering indeterministic classical mechanics we gain a second example that has similar problems and paradoxes to those found in quantum theory \cite{del2024features}. Moreover, we all have the strong intuition that we understand classical mechanics, hence conceptual problems appearing both in indeterministic classical mechanics and in quantum theory might be easiest to solve thanks to our classical intuitions.

The split of trajectories in Fig.~1.c makes it quite clear that something has to happen, nature has to make up her mind. But this is not precise enough. When does nature have to make up her mind? I do not know, especially not in the quantum domain where chaos in the usual sense of hypersensitivity to initial conditions does not exist. But allow me to present a possibility. Assume that nature can produce/create new bits of information, as assumed during quantum measurements, and assumed in naturalised intuitionistic mathematics \cite{NatInt}. But assume she can't produce arbitrarily large quantities of fresh information in a finite time. Let's denote $\gbar$ the maximum number of new bits per unit time and volume\footnote{A consequence of this assumption is that, if the Universe is finite both in time and in space, there is a kind of largest ``number", $\gbar$ times its age and its volume. Hum, not sure I take that too seriously. But why not?}, hence the units of $\gbar$ are bits$/s/m^3$. As long as the number of alternative possibilities -- the number of potentialities -- is small, $\gbar$ has no effect. But what if the system is a quantum system that is classically chaotic?

In some sense there is no quantum chaos, unitarity imposes that the scalar product of 2 possible initial conditions remains constant in time\footnote{Ballentine notices that something similar happens in Hamiltonian classical dynamics, the overlap of two probability distributions remains constant \cite{BallentinClassChaos}.}. However, it is known that if a dynamical system is classically chaotic, then the quantum analogue produces enormous amount of entanglement. As a well-known example, consider the kicked top Hamiltonian \cite{KickedTop} and assume it is made out of 20 spins $\half$ (more generally of $N$ spins $\half$). Since this is a classically chaotic system, the quantum analogue is such that the 20 spins get quickly highly entangled \cite{BilalQchaosEntangl}. Assume furthermore that the initial condition is a product state of the 20 spins, but each with some indeterminacy. As in indeterministic classical mechanics, each spin points towards a point on the Bloch sphere, but with a very small indeterminacy. This is best represented by a slightly mixed state $\rho$ s.t. $Tr(\rho^2)\approx 1-\epsilon$. Note that for $N$ similar spins the initial condition $\rho^{\otimes N}$ satisfies $Tr\big((\rho^{\otimes N})^2\big)\approx 1-N\epsilon$. Hence, the initial condition requires only a finite amount of information to represent it. However, after a short time the evolution is such that all 20 spins get highly entangled, requiring about $2^{20}$ parameters ($2^N$). Each parameter is still represented by finite information. Nevertheless, the total information has to be huge, if not, some of the $2^{20}$ parameters would be totally undefined, e.g.~with less than 1 bit per parameter \cite{Palmer}. Here again, nature has to make up her mind. But now it is not merely a binary choice. Nature would have to produce an enormous amount of information in a very short time.

Let's assume that nature is unable to produce that amount of information quickly enough. What could happen? A possibility that I like to consider seriously is that nature splits the 20 spins in two parts, each of about 10 spins ($N/2$) -- or in $k$ parts, but for simplicity let's consider the case of a split in two parts each approximatively of the same size. Now, it suffices to produce enough information for $2\cdot2^{10}$ parameters ($2\cdot2^{N/2}$), a way smaller number, possibly a number of fresh bits that nature can produce in the given time.

The above sketch is very vague, certainly not enough even to claim a model. But I believe it makes some sense and is worth developing. I like it because it is a possible beginning of an explanation of why the world does not evolve into an enormously entangled ``block". Indeed, I do not see why one should extrapolate our small laboratory experiments to the entire cosmos, personally I do not believe in concepts like ``the state-vector of the universe". I believe it is important and scientifically sound to develop serious alternatives to the ``wave function of the universe". 

If my very sketchy proposal turns out productive, then the many-worlds interpretation would be falsified. Would Gilles be disappointed? I'll ask him, of course. But I bet he will be happy to see new physics, new ideas. I believe his view on the many-worlds is more a proof of logical consistency and the Paul Raymond-Robichaud model\footnote{Which also rejects the concept of a wave function of the universe.} (see next section) a demonstration that this can be done in a rather economical way.

To conclude this section, what would be the implication for quantum computers in case nature is really limited in her capacity to produce enormous amount of entanglement in a short time? I believe the consequences could be serious. If it requires lengthy time to produce the kind of entanglement needed for efficient quantum computations, then decoherence might be even more damaging than usually thought. Of course, error correction may help to mitigate this, but that means even more entanglement, hence longer times. At the end, it is not clear who would win, the devil hides in the details.

\section{Many-worlds or relativity of indeterminacy ?}\label{RelInd}
Consider a true random number generator (TRNG). This could be a quantum one, but not necessarily. It suffice to assume that some events were not necessary before they occurred. Then, clearly, there is a time before the random number was produced and a time after. Hence, time ``created" one or several new bits of information. Here time does not merely pass, but is a real process as we developed with Flavio in \cite{creativeTime}. How does that fit in (special) relativistic space-time? 

For simplicity, assume the true random number generator is small enough to be considered as point-like and produces a single bit $a$. In the diagram shown in Fig.~2 the bit $a$ is produced at the origin of the first (left) shown light-cone. In the past the bit $a$ is indeterminate. But where is the past? According to relativity, the bit $a$ is indeterminate in the entire space-time region outside the future light-cone; $a$ has a determinate value only inside the future light-cone. Accordingly, the proposition ``$a=0$" has a truth value only inside that light-cone, outside the proposition has no truth value, illustrating the failure of the Law of the Excluded Middle. 

For concreteness, assume the true random number generator produced $a=0$. Hence, in some sense, $a=0$ is a fact. But this fact is relative. Important here is to realize that this relative fact is not relative to some agent or observer, in contrast to \cite{BruknerRelFacts,WisemanRelFacts,RovelliRelFacts}, but is relative to space-time location: $a=0$ is a fact inside the future light-cone, but it is not a fact outside.

\begin{figure}[h]
\includegraphics[width=8.7cm]{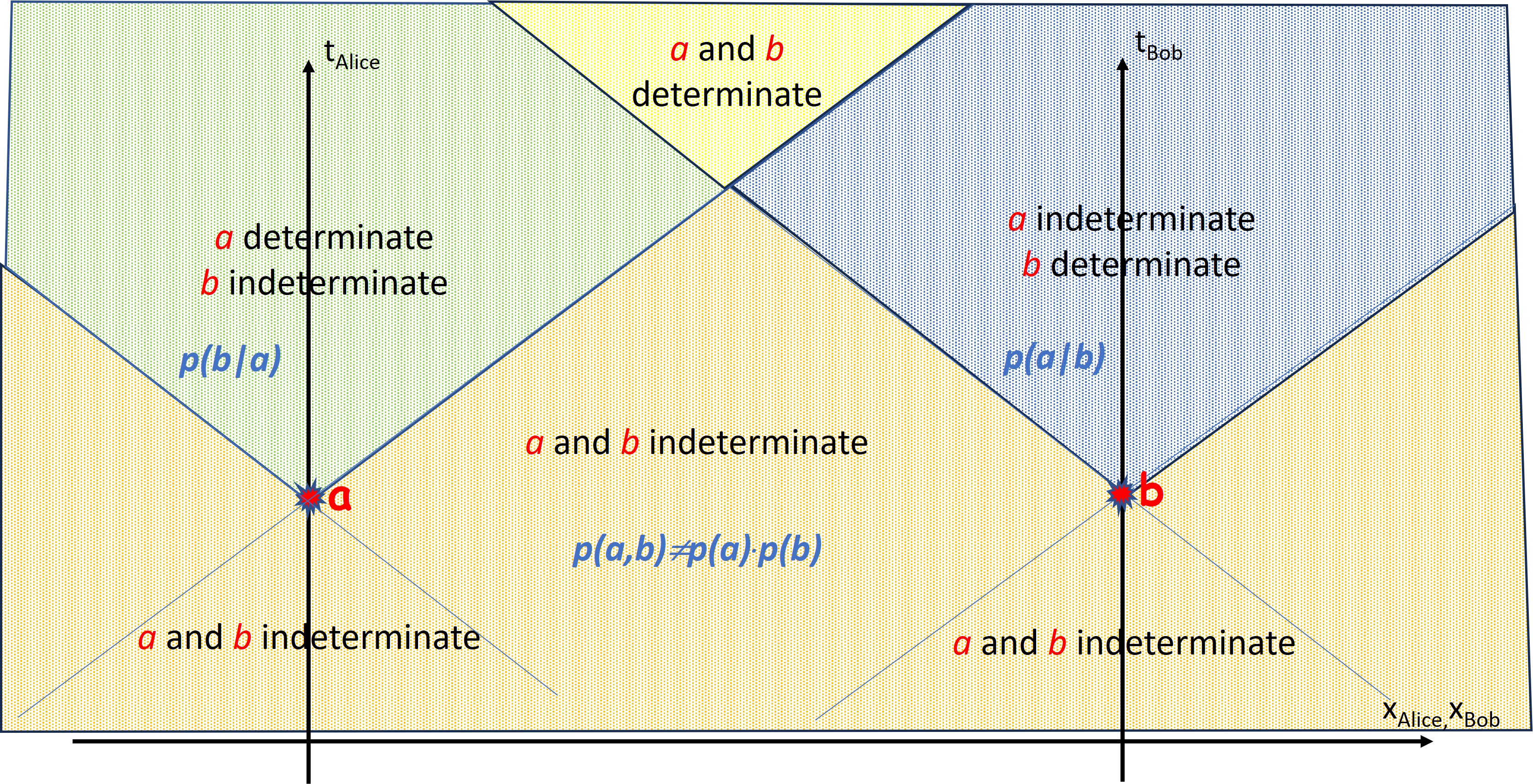}
\caption{\it Space-time diagram with two localized True Random Number Generators (TRNG) that produce bits $a$ and $b$, respectively. The first TRNG is displayed on the left, the second one on the right. Since they are space-like separated, there is no time ordering. The diagram indicates where the bits $a$ and $b$ have determinate or indeterminate values and where ontological conditional propensities (objective probabilities) are defined: $p(b|a)$ in the green region, $p(a|b)$ in the blue region and in the intersection of the two light cones - in yellow - both $a$ and $b$ are determinate.}
\end{figure} 

I believe the above picture is rather simple\footnote{Let me acknowledge that this simple picture is incompatible with classical logic. First because classical logic assumes the law of the excluded middle, hence stating that the proposition ``$a=0$" has no value is impossible in classical logic. Next, because in classical logic propositions never change value, hence since ``$a=0$" is a true proposition inside the future light-cone, it ought to have been true since ever. More work on suitable logics for indeterministic physics is needed \cite{logic4Physicists}.}. We'll soon come to the complication of quantum non-locality. But before, let us ask why the above picture is not the common one. Rietdijk and Putnam argued in the 1960's that relativity imposes determinism and, hence, that true randomness can't exist \cite{RietdijkDet,PutnamDet}. Strange. Randomness does exist. Hence, if the argument would be solid, that would, in my opinion, falsify relativity. However, the Rietdijk-Putnam argument is surprisingly weak. It assumes that the hypothetical creation of a fresh bit would be a fact along an entire space-like slice passing through the location of the TRNG. But why should anyone buy such an assumption which is clearly in strong contradiction with the entire spirit of relativity? In \cite{del2021relativity}, Flavio and I argued against the Rietdijk--Putnam argument and proposed to replace the idea that the fresh bit should ``exist" on an entire space-like slice, by the way more plausible (according to relativity!) assumption that it ``exists" only inside the future light-cone. Admittedly, I am very surprised that the Rietdijk-Putnam argument had such a large positive echo. Seemingly, scientists are particularly tolerant with proofs that support their prejudices.

But now to the real challenge: non-locality. Consider not one, but two true random number generators and assume they are not independent. Not independent even when conditioned on all their common past (past light-cone, we are still in the frame of relativity). For our purpose, the two TRNGs could be the two sides of a PR-box (such boxes do not exist, but are perfect conceptual tools to talk about non-locality), but they may also be the two sides of a bipartite entangled state, e.g.~a singlet. Notice that in such a situation it is not clear whether there are two TRNGs, or only one global TRNG, i.e.~one non-local TRNG. But in any case, there are two regions in space-time where correlated fresh bits are produced. Assume these two regions are space-like separated (Fig.~2 represents the situation seen from the reference frame in which the two events are simultaneous, since the events are space-like separated such a frame always exists). Denote $b$ the second bit and, for concreteness, assume eventually $b=1$. As above we'll assume that $b=1$ is a fact in all its future light-cone. As is well-known the two events have no time ordering. But again, both of their determinate values hold inside and only inside their respective future light-cones.

When the future light-cones meet (the yellow region on Fig.~2), both bit values are determinate and any function of them likewise. For example, $a\oplus b$ (addition modulo 2) has a determinate value only inside the intersections of the two light-cones.

How can one talk about the correlation between $a$ and $b$? When $a=0$ occurs, nothing changes at the distant location of the second TRNG. Symmetrically, when $b=1$ occurs nothing changes at the location of the first TRNG. Hence, how can $a$ and $b$ be correlated? Not from their common past (as assumed above and as proved by the violation of some Bell inequalities), not by spooky actions at a distance; physics is not spooky. Here I see two possibilities.

The first possibility assumes that, actually, $a=0$ never occurs. What really happens is that the world, locally, splits in two. In one branch $a=0$, in the other branch $a=1$. Similarly, $b=1$ happens in one local branch, while $b=0$ occurs in the other local branch. This idea is the essence of many-worlds (at least as I understand it). Then, for a while nothing happens. Until the two future light-cones merge. There, the $2\!\times\!2$ branches merge into 4 branches with ``weights" corresponding to the initial probabilities $p(a,b)$. 

I hope I did more or less justice to Gilles' view. Actually, the first time Gilles and I discussed that matter was at a workshop organised by Antoine Suarez on Science and Free-Will \cite{SuarezFreeWill}. I argued that free-will falsifies the many-world view \cite{OutsideSpaceTime,MultiversPandemic} and Gilles immediately replied (by coincidence, his talk was scheduled directly after mine) with a contribution entitled ``Can Free Will Emerge from Determinism in Quantum Theory?". In the proceedings, the idea by Paul Raymond-Robichaud of ``parallel lives" was added \cite{parallelLives,BrassardManyWorlds}. According to parallel lives, all splits happen locally, i.e.~the universe does not split instantaneously all over the places, see \cite{RaymondRobichaud1,RaymondRobichaud2}. And now, how do I see it?

In my view the two events happen locally, e.g.~$a=0$ and $b=1$, each at its own location and with one and only one outcome on each side, producing facts in their future light-cones. Recall that one event does not change anything at the other's location -- no spooky action at a distance. However, the two events may be correlated. This correlation between distant events is non-locality, a highly surprising feature of our world, a feature no one would trust if Bell inequality violations would not be a well-established empirical fact. Formally this non-locality is expressed by the non-factoring joint probability\footnote{All the probabilities appearing in this section could be conditioned on some inputs or on some measurement settings, traditionally denoted $x$ and $y$. Since this is not necessary for our story (the inputs could be fixed), we use simplified notations.}: $p(a,b)\neq p(a)\cdot p(b)$.

Another way of talking about this situation is that there are not really two events, but one global random event, one random event that manifests itself at several locations. Let me emphasize that such a global non-local random event does not allow one to communicate between the two locations at which it manifests itself. If one is ready to consider ontological indeterminism, hence the possibility that new information gets ``injected" into the world, why not accept that this new information may have consequences at several locations, as long as this does not allow for spooky actions at a distance?

So, at the end, what is ``better", a world able to split and recombine local branches with (mysterious) weights? Or a world full of indeterminacy in which things can really happen, even with non-local correlations, i.e.~with correlations that do not come from the past though they concern distant events? Gilles will go for the first alternative, I firmly stick to the second one. My main argument is simple: measurements have outcomes, one and only one outcome per measurement; and Bell non-locality is an empirical fact.

Let me elaborate a bit on my view. The standard conundrum is that the genuine correlation (genuine in the sense that it does not merely originate from the common past of the two events) suggests that the probability of $b=1$ seems to change once the fact $a=0$ is secured. But this picture is incorrect -- although it is the standard picture provided by (pre-relativistic) quantum theory! The objective probability (propensity) of $b=1$ does not change. For instance, for PR-boxes and for singlets it always equals $\half$. What changes is the conditional objective probability, when conditioned on the fact $a=0$. But this fact exists only inside the first light-cone, hence $p(b|a)$ is valid inside and $p(b)$ outside the first light-cone. Admittedly, $p(b|a)$ and $p(b)$ have different values. This might be surprising, but not more than the relativity of lengths and the possibility to signal inside but not outside light-cones. Symmetrically, $p(a|b)$ holds inside the second light-cone and $p(a)$ outside\footnote{Conditional propensities are computed as for mere epistemic probabilities: $p(a)=\sum_b p(a,b)$ and $p(b|a)=p(a,b)/p(a)$.}. 

Isn't the above quite simple? The tension that makes us have mixed feelings about the above story is, I believe, that we are not used to think of facts and of objective probabilities as relative to space-time locations, although we are very used to think that the events are located in space-time. This is why in \cite{del2021relativity} Flavio and myself named our approach ``The Relativity of Indeterminacy"; an alternative title could have been ``The Relativity of Propensities", but at the time we had not yet written our paper on propensities (rightly entitled ``Potentiality Realism" \cite{del2023prop}).

So, at the end, what is my view of non-locality, i.e.~of the violation of Bell inequalities? Here it is.
First, the local propensities (objective probabilities) never change (until they turn into local facts/events or gain new information by entering space-time regions where this new information exists). Second, the conditional propensities depend on the condition(s), hence they are as local as the events on which they are conditioned. Finally, once the light cones merge, there are no longer propensities nor conditional propensities, but facts (or the event $b$ did not yet actualize). Admittedly, it is remarkable that the observed statistics violate Bell locality, but nothing in this story is spooky: indeterminacy is relative, as are lengths and (geometric) time.

Surprisingly, the most counter-intuitive picture is when there is no actualization on Bob's side, i.e.~$b$ remains indeterminate. This happens if no input is fed into Bob's side of the PR-box or if no measurement is carried out on Bob's end of the singlet. In such a situation the propensity of $b$ remains $\half$, $p(b)=\half$, until Bob's TRNG enters Alice's future light-cone; then and only then the propensity of $b$ is conditioned on $a$: $p(b|a)$. According to this picture the propensity of $b$ changes abruptly. But notice, it changes locally. The change is abruptly like the possibility to signal from location A to location B: this possibility also changes abruptly when location B enters A's future light cone. Notice that this change of value of the propensity of $b$ can't be detected; this results from the same line of arguments as the famous impossibility to signal using entanglement. Indeed, if one could detect the change in value of the propensity of $b$ when it enters A's light-cone, then one could equally detect a change on B's side when A's qubit is measured in pre-relativistic space-time, figures 2 and 3 are very similar, except that in the later the light-cones open to half spaces.

\begin{figure}[h]
\includegraphics[width=8.7cm]{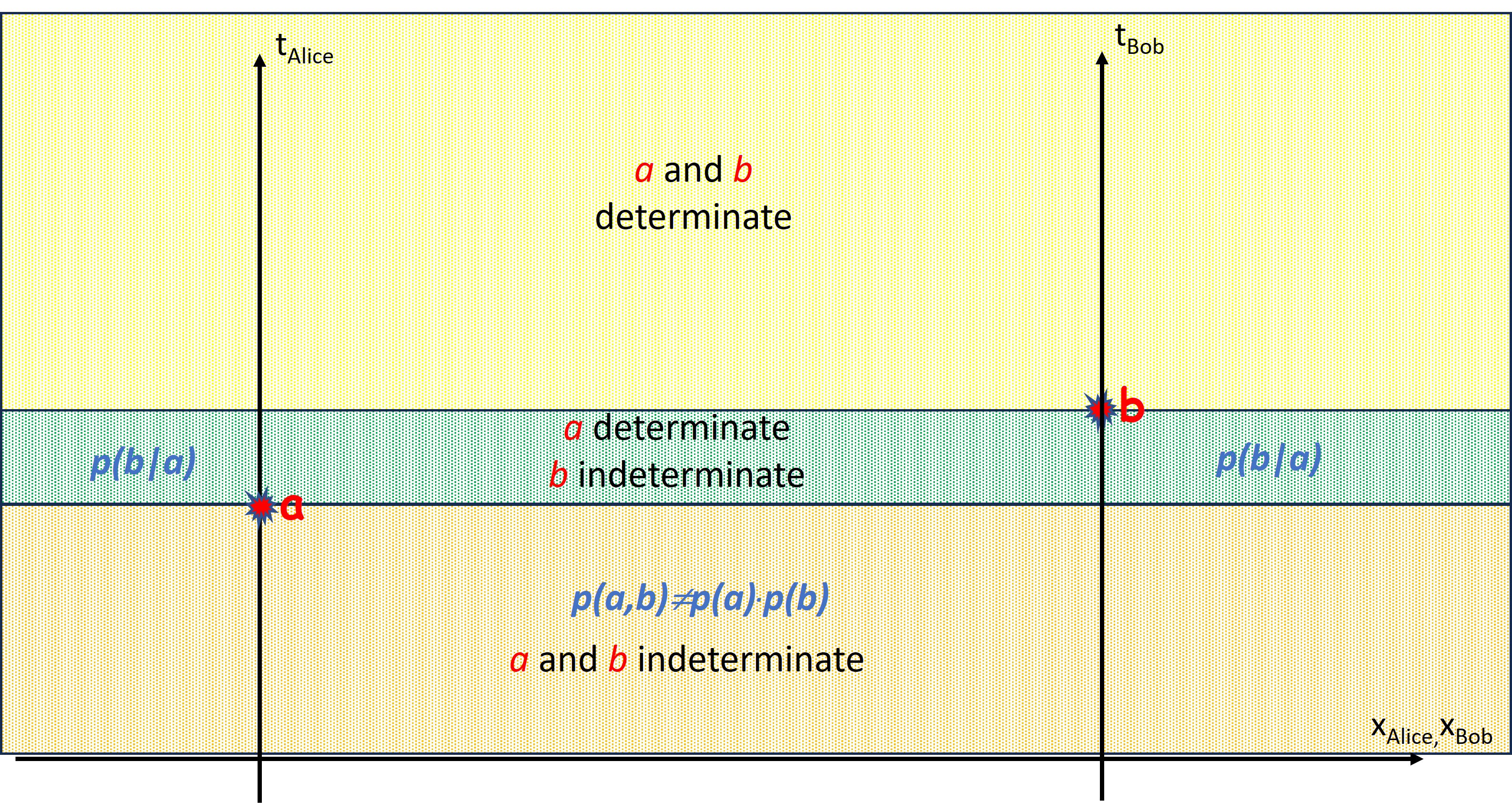}
\caption{\it Similar to Fig.~2, but in a pre-relativistic Newtonian space-time in which unbounded velocities are allowed.}
\end{figure} 

To conclude this section let me contrast the above story with the equivalent one in pre-relativity, i.e.~in Newtonian space and time with unbounded velocities and absolute simultaneity. This corresponds to widening all light-cones to half-spaces. Hence, as soon as the first TRNG produces bit $a$, the fact $a=0$ 
is a fact in the entire absolute future of the event, extending instantaneously to the location of the second TRNG, assuming it lies in the future as in Fig.~3. Hence, instantaneously, conditional probabilities like $p(b|a)$ are valid, giving the impression of a spooky action at a distance. This, in a pre-relativity context, is not surprising since unbounded velocities are assumed. What is highly surprising however, is that seemingly Einstein used to think of the EPR paradox in a pre-relativistic world-view. Can that be? I'll let historians debate about this. For me the situation is pretty simple, if unbounded velocities are allowed in the theoretical framework, then one should not be surprised by apparent ``action at a distance". But if velocities are bounded, e.g.~by the speed of light, then there is no ``action at a distance" and conditional probabilities like $p(b|a)$ make sense only where the condition $a$ exists, i.e.~in the future light-cone of the TRNG producing $a$.

The only counter-intuitive feature that remains is that the joint probability $p(a,b)$ does not factorize\footnote{Even if conditioned on possible inputs -- measurement settings -- and states, whether denoted by some $\psi$ or some $\lambda$.}: $p(a,b)\neq p(a)\cdot p(b)$. I wrote above that a way to understand that is to consider $(a,b)$ as a single de-localized event. Possibly, a better notation would be $p\big( (a\!,\!b)\big)$, illustrating that $(a\!,\!b)$ is a single object.

\section{conclusion}
Thank you Gilles for forcing me to put down the ideas expressed here. It was fun writing this. Let me emphasize the statement that facts are relative, but {\bf not} relative to some observers or agents, relative to their space-time locations, as one should expect in a relativistic context. This allows one to consider indeterminism also in relativity, including correlated indeterminism such as that which can violate a Bell inequality, without any spooky action at a distance. All the surprise resides in the non-factoring of the propensities: $p(a,b)\neq p(a)\cdot p(b)$.

The other idea expressed here is that nature might be limited in her power to produce fresh information (information that did not exist and was not necessary). Nature could have the power to create new information, but only at a limited rate. This would have dramatic effects in quantum dynamical systems producing a lot of entanglement, as in those that are classically chaotic. Here I like to emphasize that information as used in this contribution is {\bf not} information of an agent nor of an observer, but ``information of nature", possibly better expressed as structure that exists out there.

As one sees from the two previous paragraphs, agents and observers are not necessary to do physics, not more than privileged reference frames. Of course, looking for privileged reference frames may be a valid research program, but only if it leads to new physics, to new predictions \cite{Salart,Cocciaro11,Cocciaro18,China,Bancal12,GisinHiddenInfluence,Pignard19,Papatryfonos,RevBeyondBell}. Similarly, looking for observer-dependent facts should lead to new physics.

Are we near new physics or shall we remain in a state of normal science as defined by Kuhn \cite{Kuhn}. Recently a revolution came from AI (recall that Gilles is a computer scientist). I am very impressed by AI's ability to solve technical problems. But will AI ever propose a change of paradigm?

\small
\section*{Acknowledgment} 
No AI has seen this paper, nor any previous version of it. Accordingly, I take personally full responsibility for all typos and ideas in this paper.
Thanks are due to Alejandro Pozas Kerstjens, Jef Pauwels, Flavio del Santo, Sajjad Bagdadi and Bilal Khalid for useful comments.

\end{document}